# Studies of long-lived photogenerated carriers in low band gap polymer photodiodes

Monojit Bag and K. S. Narayan, *Jawaharlal Nehru Centre for Advanced Scientific Research, Jakkur P.O., Bangalore-560064, INDIA*

*Abstract*—Defects in low-bandgap polymer based photodetectors play a critical role in determining switching characteristics in the near infra-red spectral-regime relevant to communication wavelength. We carry out detailed studies of the bandwidth limiting factors of the long-live transient photocurrent at different incident wavelength as a function of temperature, light pulse width, bias voltage. The results indicate that dominant transport mechanism of the photogenerated carriers is limited by recombination at low temperature and detrapping rate at high temperature. A general trend of a slower response of photocarriers originating from the band-tail region is observed.

## I. INTRODUCTION

The advent of interesting molecular systems which have been introduced by organic solar cell research community over the last few years include semiconducting donor polymer which have absorption deep into the near infra-red (NIR) region (wavelength, $\lambda \sim$ 0.8 - 1.2 µm range or bandgap $E_g$ = 1.6 eV – 1.1 eV)[1-2]. These materials are also promising as the active element for photodetectors in this range. Photodetectors in the optical communication wavelength which corresponds to $Er^{3+}$ transitions from excited state $^4I_{13/2}$ to ground state $^4I_{15/2}$ (in the range of 1.53 µm)[3] are typically made of InGaAs, Ge, PbS and PbSe. The availability of polymer semiconductors with advantages in high optical cross-sections, sizable and ultrafast nonlinear responses, color tunability, large active areas, mechanical flexibility or low-cost processing are important. In fact, recent trends in this direction involve development of inorganic-organic hybrid systems[4-5] where the advantages of the different components are fully utilized. Organic materials are used for light source; organic light emitting diodes (OLEDs)[6-7], organic light emitting transistors [8] and organic semiconductor lasers [9-10]. Organic materials are also used as transmission media for optical communication such as polymer optical fibers [11-12]. An all optical device with polymer based light source, polymer based waveguides and polymer based photodetectors are ingredients for a complete polymer based optical circuits[13-14].

The external quantum efficiency of organic photodiodes have reached up to ~80% in the visible range[15]. Band gap of polymer semiconductors and the active spectral range are typically in the 300 – 630 nm range. To achieve high sensitivity in an organic photodetector device, a common underlying strategy implemented is the utilization of the bulk heterojunction (BHJ) concept consisting of an interpenetrating network of electron donor (D) and acceptor (A) materials. Photoexcitation in the vicinity of BHJ results in a charge transfer process between the D-A species within an ultrafast timescale. The charge carrier transport occurs through the network of the D-A to the respective electrodes and the device operated in a reverse bias mode exhibits large responsivity. The pathway to low-bandgap semiconducting polymers with deeper HOMO levels have been demonstrated[16-18], and BHJ solar cells fabricated by these D-A polymers have led to strategies which do not compromise on the open circuit voltage ($V_{OC}$). Additionally charge transfer (CT) states also provides source for significant photocurrent in NIR regime[19].

Unlike the explosive activity in organic BHJ based solar cells, results from NIR based photodetectors are sparse[20-21] and includes reports of high detectivity polymer photodiode in a wide spectral range (300 nm -1450 nm)[22]. Apart from the spectral requirement, switching speed and noise equivalent power (NEP) are also important parameters, which we emphasize in this report. We highlight the dependence of these parameters on temperature ($T$), pulse width ($T_{ON}$), intensity, wavelength ($\lambda$) and background dc light conditions. We also demonstrate the difference of transient photocurrent [$I_{ph}(t)$] response originating from the charge transfer complex (CTC) and the photogenerated carriers from the direct excitation of donor polymer.

The early processes involved in photoinduced charge carrier generation and separation is quite rapid (in sub-picoseconds regime) in these systems[23]. Modification of polymer-BHJ devices with additives have recently shown to increase the carrier mobility, generation rate as well as free carrier lifetime. The expected $I_{ph}$ switching response of these photodiodes should be in the range of nanosecond scale. However, observed transient response from a BHJ photodiode indicate a

This work was supported by Department of Science and Technology, Govt. of India for the Indo-UK project on excitonic solar cells.

K. S. Narayan is with Jawaharlal Nehru Centre for Advanced Scientific
K. S. Narayan is with Jawaharlal Nehru Centre for Advanced Scientific Research, Jakkur P.O., Bangalore-560064, INDIA (e-mail: narayan@jncasr.ac.in ).

M. Bag was with Jawaharlal Nehru Centre for Advanced Scientific Research, Jakkur P.O., Bangalore-560064, INDIA (e-mail: monojit@jncasr.ac.in ).

slow decay with a presence of $I_{ph}(t)$ even at millisecond range[24]. The long-lived $I_{ph}(t)$ is attributed to the presence of large number of trap states in these systems. Transient photocurrent (TPC) studies have been carried out by many groups to address carrier transport model in these disorder semiconductor structures[25-27]. Drift-diffusion models have been invoked to explain the slow decay of $I_{ph}(t)$. The effect of trapping and de-trapping of charge carriers and the recombination kinetics have been addressed in many of their papers at different bias voltages, different intensity of probe light or in presence of background light[24]. We emphasize the $T$ dependent $I_{ph}(t)$ measurement and address the nature of the trap density. Apart from the explanation of the slow decay in terms of trap kinetics, interpretation in terms of photoinduced gap states are also common. In the well studied model system of P3HT:PCBM, free carrier lifetime ($\tau_e$, ~ 300 - 400 ns)[29] strongly depends on the carrier concentration ($n$) and bimolecular recombination rate ($\beta$)[28]. As $n$ increases, $\tau_e$ decreases by a factor of $n^{-\delta}$, where $\delta$ lies within 0.95 to 1.05. $\beta$ increases with the increase in carrier mobility ($\mu$), especially the slower carrier species in a BHJ devices. It is expected that upon light termination, $I_{ph}(t)$ should rapidly decay back to dark current magnitude in absence of any charge retention. The electron mobility ($\mu_e$) is higher than hole mobility ($\mu_h$) in P3HT:PCBM system, hence $\beta$ is defined by $q\mu_h/\varepsilon_r$, where $\varepsilon_r$ is dielectric constant of the blend material. High mobility of charge carrier may result in high recombination loss. The shorter carreier lifetime then cannot explain the slow decay processes observed in form of long-lived $I_{ph}(t)$.

The ground state interaction between donor-HOMO level to acceptor-LUMO level results in the form of a CTC which have lower energy than individual bandgap energy of each material. Direct excitation of CTC (polaronic absorption) may result in slow dissociation of charge carriers due to low thermal energy (effective free energy for polaron pair dissociation)[29]. Those low energy carriers may get trapped in the bulk and result in slower recombination rate. However, Lee et. al. recently demonstrated that the internal quantum yields of carrier photogeneration are similar for both, excitons and direct excitation of CTC[30]. The formation of CTC is a function of donor-acceptor ratio, their relative band position and processing condition like solvent assisted and thermal annealing[31].

## II. EXPERIMENT

Low bandgap donor type polymer poly [N-9'- heptadecanyl-2,7-carbazole-alt-5,5-(4',7'-di-2-thienyl-2',1',3'-benzothiadiazole)] (PCPDTBT)[32] was blended with the acceptor materials such as [6,6]-phenyl-$C_{61}$-butyric acid methyl ester (PCBM) or $C_{70}$ based PCBM (1:3 by weight ratio) in chlorobenzene solvent (20 mg/ml concentration) and spin coated inside glove box at 1000 rpm speed for 60 second on Poly[3,4-ethylenedioxythiophene]:poly[styrenesulfonate] (PEDOT:PSS) coated ITO substrates to get ~100 nm thick film[33]. 100 nm thick Al cathode was thermally deposited using a physical mask (1 to 10 $mm^2$ active area) at $10^{-6}$ mbar pressure. The results were verified by testing several devices. $T$ dependence $I_{ph}(t)$ at zero bias was carried out using a liquid He based cryostat chamber. 532 nm pulsed laser (10 ns pulse) with a repetition rate at 10 kHz was used as a light source [Fig 1.(a)]. Wavelength dependent transient measurement is carried out using 470 nm (blue) LED and 930 nm IR LED. Pulse width, bias voltage and intensity dependent transient photocurrent measurements were carried out using 690 nm ultra-bright red LED. $T_{ON}$ was controlled by controlling the pulsed input voltage profile from a function generator (Tektronix AFG320). A white LED (GaN based) is used as a background light source. All transient measurement was recorded by a GHz LeCroy oscilloscope (Waverunner 6100A). Transient photocurrent was recorded from ITO as anode electrode while Al cathode was grounded. A low noise, low impedance pre-amp (current to voltage with a gain $10^5$ V/A, 50 $\Omega$ coupling with a bandwidth of 400 kHz) is used to avoid loading to the sample and associated time delay due to external resistance [Fig 1(b)]. The spectral responsivity from 400 nm to 900 nm was tested using a monochromator (Spectromax 500M), Xe lamp source and a Si–photodiode for calibration [Fig 1(c)].

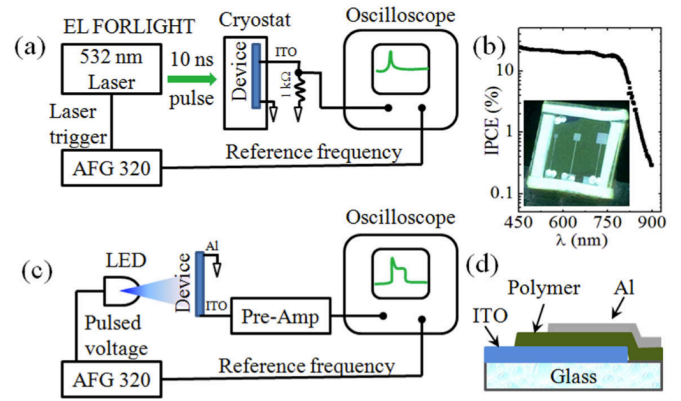

Fig. 1. (a) Schematic diagram of low temperature transient photocurrent measurement setup. (b) IPCE spectrum of PCPDTBT:PCBM sample. Inset: Photo of an encapsulated device. (c) Schematic diagram pulse width dependent measurement setup. (d) Schematic cross sectional view of a device.

## III. RESULT AND ANALYSIS

The response of the device to a short single pulse of photoexcitation at different $T$ at zero bias can be expressed in a general stretched exponential form; $I_{ph}(t,T)=I_{ph}^{max}(T)\exp(-t/\tau)^\gamma$, where $\tau$ is a characteristics decay time constant and dispersion parameter $\gamma$ in the range 0 - 1[34-35]. The observed behavior of $I_{ph}^{max}(T)$ appears to be an activated type as indicated by the $\log(I_{ph}^{max})$ Vs. $T^{-1}$ response in figure 2(a), with activation barrier of $\varepsilon_a$ ~16 meV. The temporal part of the photocurrent decay can be reasonably modeled to a simple exponential function ($\gamma \approx 1$) $I_{ph}(t)=I_{ph}^{max}\exp(-t/\tau)$ in the temperature range 200 – 300 K. It is to be noted that the coefficient $I_{ph}^{max}$ depends on the light intensity, carrier mobility-lifetime product and the density of trap states[36]. It is known that $\tau(T)$ in disordered systems increases with the decreasing $T$ and can be attributed to an carrier emission rate ($\propto 1/\tau$) and exponentially varies with $T$ i.e. $\tau(T) \propto \exp(\Delta/k_BT)$ where $\Delta$ represents the energy difference between transport level ($E_t$) to the site energy ($E_i$) as shown in the figure 2(b).

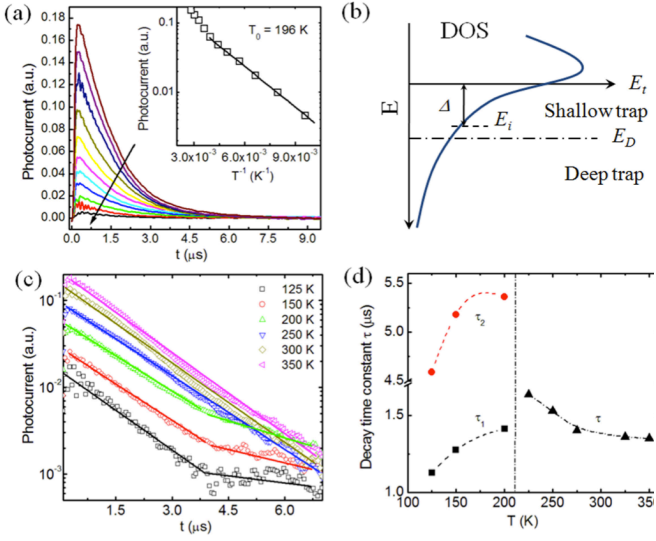

Fig. 2. (a) Transient photocurrent at different temperature. Inset: $\log(I_{ph}^{max})$ vs $T^{-1}$ plot. (b) Schematic diagram of DOS, transport energy level $E_t$, demarcation energy $E_D$ and the $i$'th site energy $E_i$. (c) $\log(I)$ vs. $t$ plot at different $T$. Above 200 K, transient photocurrent (with 1 kΩ coupling resistance) follows single exponential decay. Below this temperature the response shows multi-exponential decay. (d) Single exponential decay time constant for temperature above 200 K and bi-exponential fitting for a temperature range below 200 K.

The dynamics represented by $\tau(T)$ estimated at different $T$ regimes in the present case of a standard representative device indicates specific features. As shown in Fig. 1(c) $\tau$ of ~ 1.3 μs at $T$ ~ 350 K, marginally increases to ~1.7 μs upon decreasing $T$ to 200 K. In the regime $T$ < 200 K $I_{ph}(t)$ does not yield good fits to a single time constant. The observed transient $I_{ph}(t)$ takes on a bi-exponential decay with an appearance of an additional distinct slow time constant [Fig. 2(c)]. Both the components ($\tau_1$) and ($\tau_2$) continue to decrease upon lowering $T$, as indicated by the magnitude of $\tau_1$ ~ 1 μs, and $\tau_2$ ~ 4.7 μs at $T \approx 125$ K. The origin of this anomalous behavior of $I_{ph}(t)$ response can be attributed to the existence of $T$ dependent different trap-mediated transport mechanisms operative in the two regimes. The occupied trap states can be classified as shallow and deep trap depending on the either side of the demarcation energy ($E_D$) level. At lower temperature ($T < T_0$), $E_D$ decreases[28] where the density of localized states near $E_t$ contributes to the transport and recombination dominates over detrapping of charge carrier. As temperature decreases, $\varepsilon_a$ also decreases as it follows $T^{3/4}$ dependence[37]. However, above a certain temperature ($T > T_0$), all occupied trap states due to photogenerated carriers are thermally accessible and hence $\varepsilon_a$ is constant in this temperature regime. Detrapping rate increases with the increasing temperature hence $\tau$ decreases. If the transient dependence can be profiled by $\tau(T)$ then at $T$ = 200 K, a crossover in the $\tau(T)$ is observed where both mechanism could be present to some extent giving rise to slowest response of the device as shown in the figure 2(d).

The manner in which the photocurrent ($I_{ph}^{rise}$) builds up to reach the equilibrium steady state value is also quite informative. The recombination processes are expected to be dominant factor in the $I_{ph}(t)$ rise-profile description. The relevant time constants for these processes can be obtained by studying $I_{ph}^{rise}$ ($t \leq T_{ON}$) upon varying the pulse width of the photoexcitation source. For instance, the results from device illuminated by 690 nm LED driven by a voltage pulse of 1% duty cycle with a repetition rate varying from 1 kHz to 1 Hz are shown in figure 3(a). The transient response when light pulse is ON can be viewed from a perspective of accumulation of charges in filling the trap states along with recombination processes. Similar response was observed when incident light intensity was varied[24].

Upon switching OFF the source, a slow emission of charge carrier occurs with the rate, which depends on the occupied density of trap states ($n_t$) and/or the initial carrier density (i.e. the duration of ON pulse). The extracted charge carrier density [$n_t = \kappa (T_{ON})^\alpha$, $T_{ON}$ is in μs] follows a sub-linear dependence of light pulse ($\alpha = 0.73$) as shown in the figure 3(b). A simplified model based on the dynamics of trapping, detrapping and recombination processes can be used to explain the light pulse dependent response. A localized state at an energy $\varepsilon$ measured from $E_t$ is characterized a density of states $M(\varepsilon)$ and the occupied density of state is $m(\varepsilon)$[38]. If the generation rate is $n_i(t)$, the rate equation for the occupied trap density of state as given in the equation below

$$\frac{dm(\varepsilon)}{dt} = n_i(t)\left[M(\varepsilon) - m(\varepsilon)\right] v\sigma - N_c v\sigma\, m(\varepsilon)\exp(-\varepsilon/k_BT) \quad (1)$$

Here $v$ is the thermal velocity of the carriers $\sigma$ is the capture cross section and $N_c$ is the effective density of state in the conduction band. The solution of $m(\varepsilon)$ then takes a exponential decay form with a time constant that depends on $n_i(t)$, $v$ and $\sigma$. The carrier transport rate equation is given by

$$\frac{dn(t)}{dt} = n_i(t) - \int_0^{E_D} \frac{dm(\varepsilon)}{dt}d\varepsilon - \frac{n(t) - n_d}{\tau_R}$$
(2)

where $n_d$ is the dark carrier density, $\tau_R$ is the recombination rate. $E_D$ represents the carrier demarcation level above which extended band-states interacts with the localized states. In the present case, the generation rate $n_i(t) = n_0[U(t) - U(t - T_{ON})]$ where $U(t)$ is a unit-step function. The first term in the equation 1 represents the trapping rate and appears when photoexcitation is present, and second term represents the detrapping of the charge carriers which largely shapes the decay profiles. The recombination rate (3rd term in the equation 2) on the other hand explains the features of $I_{ph}^{rise}$ ($t \leq T_{ON}$) as function of pulse width and intensity. The loss of trapped charge carrier due to recombination gives rise to sub-linear dependence of extracted $I_{ph}(t)$ over a range of 50 μs to 10 ms ON time. The solution of this expression for transient photocurrent originating from a pulsed light source is shown in figure 3(a) inset.

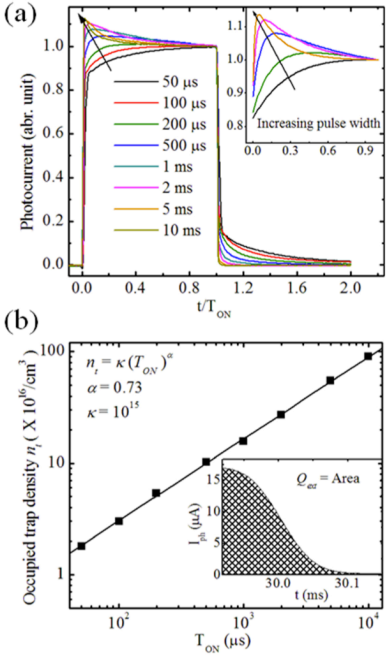

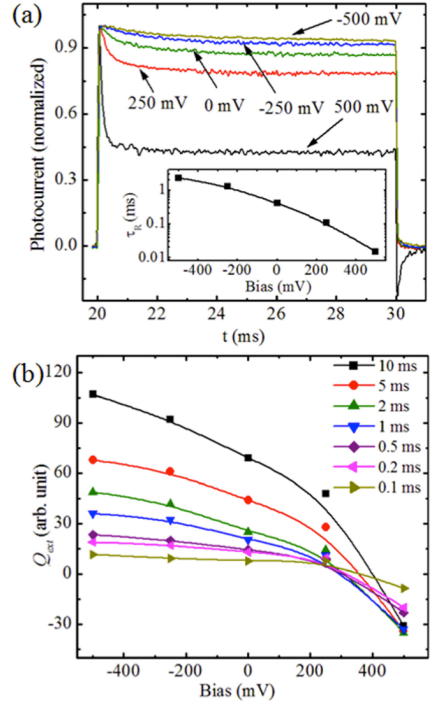

Fig. 3. (a) Room temperature transient photocurrent at zero bias (50 Ω coupling resistance) measurement with $T_{ON}$ varying from 50 μs to 10 ms. Normalized (at $t/T_{ON} = 1$) $I_{ph}$ is plotted with respect to normalized time ($t/T_{ON}$). Inset: transient photocurrent for the $T_{ON}$ period estimated from the equation 2. (b) The occupied density of trap states ($n_t$) are plotted with respect to the $T_{ON}$ time. Inset: Integrated area under the curve after the light pulse is OFF represent the total number of trapped charge carriers extracted due to thermal detrapping.

Fig. 4. (a) Transient $I_{ph}$ at room temperature for a 10 ms pulse width at different bias voltage. Inset: Recombination rate as a function of bias voltage. (b) Extracted charge carrier after light pulse is OFF as a function of bias voltage for different $T_{ON}$ varying from 100 μs to 10 ms.

The recombination loss factor is also dependent on the biasing voltage. The bias dependent photocurrent transient is depicted in figure 4(a). At $V_{OC}$, recombination loss is maximum, where as with the increasing reverse bias it can be reduced significantly. The stored charge with a longer exposure time ($T_{ON}$) in the device also shows a strong bias dependence at $V_{OC}$, [Fig. 4(b)] whereas with the decreasing pulse width, $I_{ph}(t)$ and the magnitude of the stored-charge becomes bias independent.

The effect of microscopic processes on the device switching parameters is evident in the λ-dependent studies. It is observed that the response time of the device to blue light (470 nm, ~ 1 mW) pulse is clearly faster than the NIR light (930 nm, order of μW) pulse [Fig. 5(a)]. The slower and lower response to NIR light can be attributed to the fact that the photogenerated carriers have contributions originating from the donor-acceptor CTC. Also at low carrier concentration as in the case of NIR light-pulse photoexcitation, the average mobility gets reduced by a power law behavior of $n$ where, $\mu \propto n^{(T_0/T)-1}$, where $T_0$ is related to the width of the Gaussian density of states[39]. In this case the density of occupied states is deep (high activation energy), and associates the slow carriers with long transit times and slow build-up of $I_{ph}^{rise}$ ($t \leq T_{ON}$).

The device switching response as expected is dependent on the background CW illumination. The CW white background typically results in device operation at higher charge carrier concentration. Under this condition, despite a decrease in transient photocurrent magnitude the switching response is enhanced as the rise and decay time constant decreases. The excess photogenerated carrier at steady state (space charge limited regime) reduces the exciton dissociation rate due to reduced built-in voltage and also increases the bimolecular recombination contribution. Hence in presence of a CW white background light source (same as that of a probe intensity), the $I_{ph}(t)$ decay rate significantly improves as indicated by the magnitude of τ of ~ 0.6 ms compared to the device in absence of background illumination (~2 ms) [Fig. 5(b)]. This indicates the slow release of trapped charge carriers after switching off the pulsed illumination, which has a smaller contribution as the states are completely filled due to CW background light. Under CW light, the kinetics essentially can be described dynamic equilibrium condition, i.e. the trapping and detrapping rates are almost balanced. These results are consistent with McNeill et. al. who have also shown that increasing CW background light can reduce the decay time constant[26].

IV. CONCLUSION

Low band gap polymer PCPDTBT-PCBM blend based BHJ devices photodetector device can exhibit a wide spectral response extending to 1 eV (~ 1200 nm), with a maximum responsivity of 150 mA/W at 770 nm. The trap assisted electrical transport arising from carrier generation in the tail states results in a slow extraction of charge carrier which leads to a long-lived response in the $I_{ph}(t)$ measurements. Fast rise and fast decay of transient photocurrent are important attributes for high speed detector application in the NIR regime, which in the present case can be achieved by lowering the temperature below critical value or applying a reverse bias. The reverse bias mode of operation and background-dc intensity improves the switching response. The understanding of origin of defect states and energetic along with a control of the distribution by different processing conditions can further optimize the device.

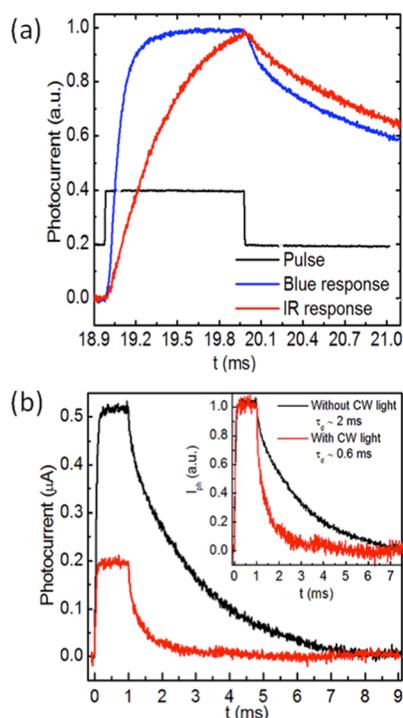

Fig. 5. (a) Transient photocurrent at zero bias (with 1 MΩ coupling resistance) response at room temperature from blue LED (blue line) and IR LED (red line) with a pulse (black line for both light sources) of 1 ms. 470 nm blue LED and 930 nm IR LED is used with a pulsed voltage source (10 % duty cycle at 100 Hz frequency). (b) Transient measurement with a blue LED (470 nm, ~ 1 mW/cm$^2$) was carried out with and without background white light at zero bias. Inset: normalized response for transient photocurrent under dark background as well as CW white background illumination.

**K. S. Narayan** obtained MSc. Physics from IIT (Bombay), and Ph.D. from - The Ohio State University. KSN's current research activities is focussed on organic electronic devices. He has over 80 publications and couple of International Patents. He is an Editorial Board Member of Pramana Journal of Physics, and a Senior IEEE member. He is a Fellow of National Academy of Sciences, India and Fellow of Indian Academy of Sciences

**M. Bag** obtained MSc. (Physics) from University of Pune in 2006 and Ph.D (material science) from JNCASR, Bangalore in 2011. His area of reach is focused on bulk-hetero-junction polymer devices, fabrication and electrical characterization. He has currently joint as Research Associates in UMass, Amherst.